# Quantification of atmospheric carbon dioxide from the Geostationary Operational Environmental Satellite (GOES East)

Aaron Sonabend-W[1], Sean Campbell[2], John Platt[1], Christopher Van Arsdale[1], Anna M. Michalak[1,3]

[1]Google Research, [2]Google Ads, [3]Carnegie Institution for Science

## Abstract

There is a growing urgency to track greenhouse gasses with the resolution, precision and accuracy needed to support independent verification of $CO_2$ fluxes at local to global scales. The current generation of space-based sensors, however, only provides sparse observations in space and time. This challenge has fueled interest in the potential use of data from existing missions originally developed for other applications to infer global greenhouse gas variability. The Advanced Baseline Imager (ABI) onboard the Geostationary Operational Environmental Satellite (GOES-East), operational since 2017, provides full coverage of much of the western hemisphere at 10-minute intervals from geostationary orbit across 16 spectral channels at an approximately 2 $km^2$ spatial resolution. Here, we leverage this high spatial coverage and temporal revisit to develop Deep$XCO_2$, a single-pixel, physics-guided neural network to estimate dry-air column $CO_2$ mole fraction ($XCO_2$). Deep$XCO_2$ employs a time series of GOES-East's 16 spectral bands, ECMWF ERA5 lower tropospheric meteorology, MODIS surface reflectance, solar and satellite viewing geometry, and day of year. The network was trained on collocated GOES-East and OCO-2/OCO-3 observations. Deep$XCO_2$ is able to capture realistic $XCO_2$ variability when compared against a held-out year of OCO-2 and OCO-3 observations, and against observations from the TCCON network. We also present case studies illustrating the use of Deep$XCO_2$ to observe $XCO_2$ enhancements over urban areas and drawdown over agricultural regions.  Overall, while the precision of GOES-East derived $XCO_2$ can never rival that of dedicated instruments, the unprecedented combination of contiguous geographic coverage, 10-minute temporal frequency, and multi-year record offers the potential to observe aspects of atmospheric $CO_2$ variability currently unseen from space.

## Introduction

Accurate characterization of atmospheric carbon dioxide ($CO_2$) variability is essential for understanding the global carbon cycle and for providing the independent verification of greenhouse gas fluxes needed to assess progress toward national and international climate commitments (e.g., Ciais et al., 2015). Carbon cycle science requires observations that resolve the spatial and temporal scales of variability of biospheric, oceanic, and anthropogenic $CO_2$ exchanges, from the diurnal cycle of photosynthesis and respiration to the seasonal drawdown of $CO_2$ across continental ecosystems (e.g., Hammerling et al., 2012). Current space-based observations remain too sparse in space and time to fully meet these needs.

The current generation of dedicated $CO_2$ satellites has transformed carbon cycle science but faces fundamental constraints from orbital mechanics and instrument design. NASA's Orbiting Carbon Observatory-2 (OCO-2), launched in 2014, retrieves column-averaged dry-air mole fractions of $CO_2$ ($XCO_2$) in a narrow ~10 km swath with a 16-day repeat cycle, achieving single-sounding precisions of approximately 1–2 ppm with a local overpass time of 1:30 pm (Crisp et al., 2004; Eldering et al., 2017). OCO-3, launched in 2019 and installed on the International Space Station, samples a range of local times and includes a Snapshot Area Map mode for targeted observations (Eldering et al., 2019). The spatial resolution of observations from both instruments is around 2 $km^2$. JAXA's Greenhouse Gases Observing Satellite (GOSAT) and GOSAT-2, launched in 2009 and 2018, respectively, achieve a repeat cycle of approximately three days, but at the cost of very sparse spatial sampling and a coarser spatial footprint of around 80 $km^2$ (Kuze et al., 2009; Suto et al., 2021). Cloud contamination also limits coverage across all passive solar backscatter missions, with valid data yields from OCO-2 in the humid tropics falling below 1% (Frankenberg et al., 2024).

Two missions launched in 2025, GOSAT-GW and MicroCarb, will augment these observations, but neither is yet producing routine science-quality $XCO_2$ products. The most significant near-term improvement is likely to come from the European Space Agency's Copernicus Anthropogenic $CO_2$ Monitoring mission (CO2M), a planned constellation of three satellites designed to achieve $XCO_2$ precision of 0.7 ppm at 2 × 2 km resolution over a 250 km swath, enabling near-daily global coverage and detection of emissions at the scale of individual facilities and cities (Meijer et al., 2020; Janssens-Maenhout et al., 2020). The first CO2M satellite is not expected to launch before late 2027 (EUMETSAT, 2025), however. Overall, the network of CO2-observing satellites will continue to be sparse for the foreseeable future, and even as the network expands we will be limited to using instruments currently in orbit for understanding historical patterns.

This challenge has driven interest in extracting greenhouse gas information from instruments not originally designed for this purpose. Early examples involved thermal infrared meteorological sounders. Chahine et al. (2008) and Maddy et al. (2008) showed that the Atmospheric Infrared Sounder (AIRS) contained recoverable information about mid-tropospheric $CO_2$. Kulawik et al. (2010) extended this approach to the Tropospheric Emission Spectrometer (TES) on the Aura spacecraft, originally designed for tropospheric ozone, demonstrating retrieval of lower-tropospheric $CO_2$ profiles. ESA's Infrared Atmospheric Sounding Interferometer (IASI) on MetOp meteorological satellites was similarly shown to contain retrievable upper-tropospheric $CO_2$ information (Crevoisier et al., 2009). In the shortwave infrared, the SCIAMACHY instrument on ENVISAT, designed for broad global trace gas mapping, provided the first decade-long global $XCO_2$ record with near-surface sensitivity, despite single-sounding precisions of 3–8 ppm that required extensive averaging (Buchwitz et al., 2005).

More recent work has extended this paradigm to a broader range of instruments, primarily for methane point source detection. Varon et al. (2021) demonstrated that the Sentinel-2 MultiSpectral Instrument, a land-surface monitoring mission, could detect and quantify large $CH_4$ point sources. Ehret et al. (2022) applied a similar approach globally to Sentinel-2 and Landsat-8, and Turner (2024) used historical Landsat-5 imagery to detect $CH_4$ plumes in Turkmenistan from 1986 to 2011, extending the approach to data collected decades before greenhouse gas monitoring was contemplated. Thorpe et al. (2023) demonstrated detection and attribution of individual $CH_4$ and $CO_2$ point sources from NASA's EMIT imaging spectrometer. Cusworth et al. (2021) showed that AVIRIS-NG and the PRISMA satellite could quantify $CO_2$ emissions from power plants.

These point-source imaging approaches differ fundamentally from the global-scale $XCO_2$ mapping needed to constrain ecosystem fluxes and evaluate national inventories, however. Point-source detection targets localized enhancements of tens to thousands of ppm-m, while large-scale $XCO_2$ mapping requires resolving variability of order 1–10 ppm across gradients spanning thousands of kilometers.

The Advanced Baseline Imager (ABI) aboard GOES-East (GOES-16), operational since 2017, represents a compelling potential resource for atmospheric $CO_2$. Stationed at 75.2°W in geostationary orbit, GOES-East provides continuous full-disk coverage of the Americas every 10 minutes in 16 spectral bands at nadir resolutions of 0.5–2 km (Schmit et al., 2017). Band 16 (13.3 μm) is formally designated the "$CO_2$ longwave infrared band," and Band 5 (1.6 μm) falls within the same $CO_2$ absorption feature exploited by OCO-2 and OCO-3, but notably both are broadband channels. The only prior application of GOES to greenhouse gas monitoring is the detection of large $CH_4$ point sources by Watine-Guiu et al. (2023).

Here, we develop an approach to quantify atmospheric $XCO_2$ from GOES-East observations across the Americas by leveraging a single-pixel, physics-guided neural network that ingests a time series of GOES-East ABI radiances across all 16 spectral bands, augmented with ECMWF ERA5 meteorological state along the viewing path, MODIS surface reflectance, solar and satellite viewing geometry, year, and day of year. The network is trained on approximately 3 million collocated GOES-East and OCO-2/OCO-3 soundings from 2017 to 2023; the year 2024 is reserved entirely for evaluation against held-out OCO-2 and OCO-3 observations and coincident measurements at Total Carbon Column Observing Network (TCCON) stations across North America. We assess performance against known $XCO_2$gradients in space and time and against observed variability at TCCON stations.

# Methods

## Dataset Preparation

To estimate $XCO_2$, we developed a framework that fuses multispectral GOES-East radiances with meteorological and surface albedo data. The fused multi-instrument dataset spans from 2017 to 2024, with the final year held out entirely to assess the model's predictive skill across unseen seasonal and spatial gradients. The GOES-16 coverage used as well as the TCCON tower locations can be seen in Figure 1.

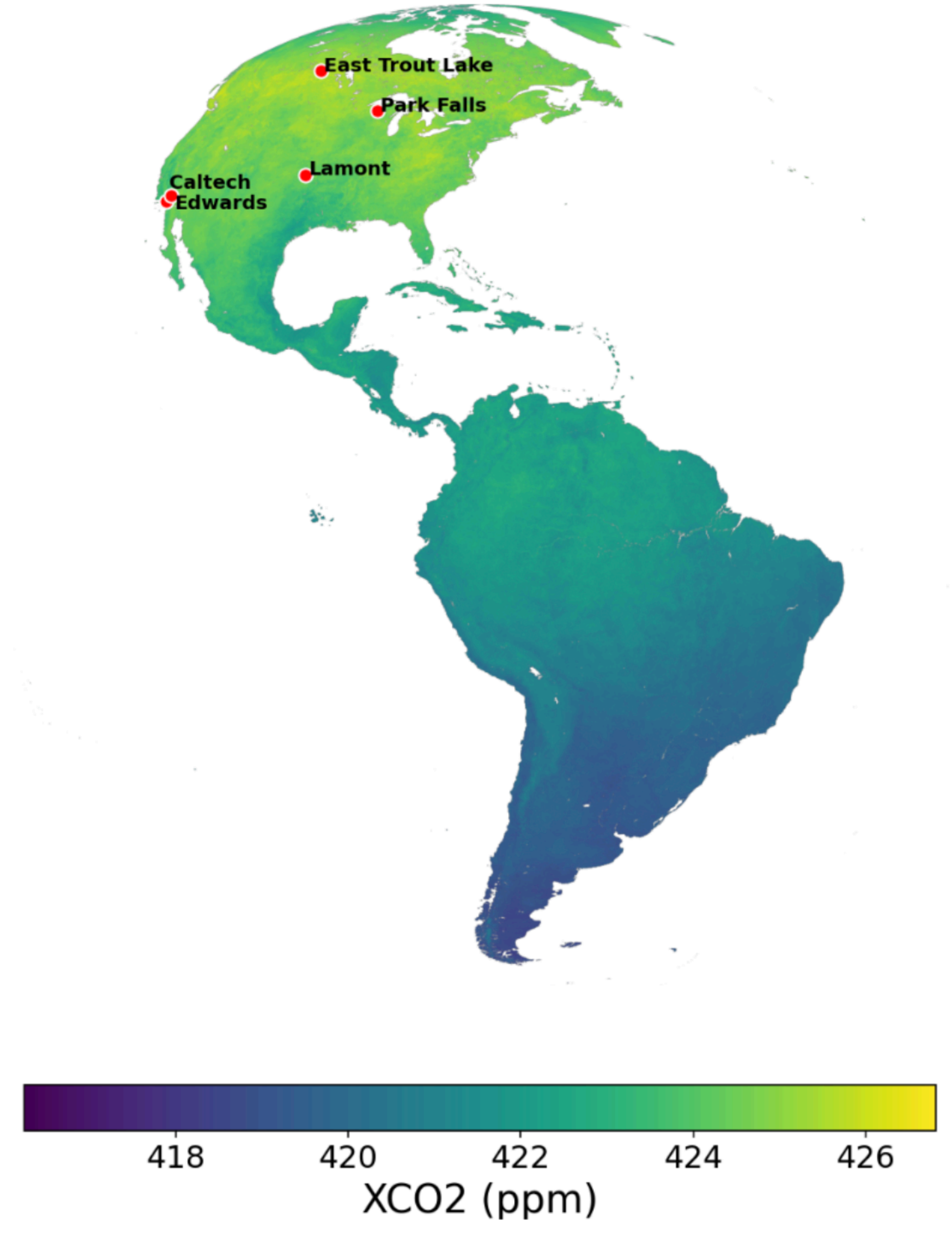

Figure 1: April 2024 average clear-sky column-averaged dry-air mole fraction of carbon dioxide ($XCO_2$) across the GOES-East footprint considered for the work, as estimated by Deep$XCO_2$. The spatial distribution captures expected latitudinal gradients and seasonal biospheric dynamics. Red markers indicate the locations of the five Total Carbon Column Observing Network (TCCON) validation towers (Caltech, Edwards, Lamont, Park Falls, and East Trout Lake) used as independent ground-truth evaluation sites in this study.

## Orbiting Carbon Observatory labels

Reference $XCO_2$ targets for model training and validation were derived from OCO-2 (v11.2r) and OCO-3 (v11r) Level 2 Lite retrieval products (Science Computing Facility, 2017, 2023). To ensure high-fidelity training labels, we selected only "best-quality" soundings (quality flag = 0), effectively excluding pixels with cloud or aerosol contamination. The dataset was further restricted to land-nadir geometries (sensor zenith angle < 2°) within the GOES-East field of view. To mitigate limb-darkening and parallax errors, we excluded observations exceeding 7,258 km from the sub-satellite point (Joyce et al., 2001). Data collocation was achieved by mapping OCO-2 and OCO-3 soundings to the nearest Advanced Baseline Imager (ABI) fixed-grid centroid (L1b full-disk coordinates) and synchronizing them with GOES-East full-disk scans within a strict 10-minute window to minimize scene drift. This collocation yielded a robust training ensemble of 3.1 million OCO-2 and OCO-3 soundings. Model performance was evaluated against a 1.6-million-pixel satellite test set.

## GOES-East observations

The model's primary features are broadband multispectral radiances from the GOES-East ABI (Schmit et al., 2017; GOES-R Algorithm Working Group, 2017). We used all 16 ABI channels, spanning the visible (0.47–0.64 μm), near-infrared (0.86–2.2 μm), and thermal infrared (3.9–13.3 μm) spectra. Clear-sky conditions were enforced using Level 2 products, restricting the analysis to high-quality, 'confidently clear' pixels via the 4-Level Cloud Mask and Data Quality Flags. Geometric metadata, including solar zenith angle (SZA) and satellite viewing angle, were extracted to calculate the Air Mass Factor (AMF) and correct for optical path length variations. While all bands contribute to the spectral signature, Band 5 (1.61 μm) and Band 16 (13.3 μm) provide the primary physical basis for $XCO_2$ retrieval: Band 5 targets surface-reflected solar radiation sensitive to total column and near-surface variations, while Band 16 captures thermal emissions modulated by upper-tropospheric $CO_2$ absorption.

To partially compensate for the ABI's coarse spectral resolution, we provided the model with a 100-minute temporal sequence (ten scans from $T_0$ to $T_{-90}$ minutes). This window enables the model architecture to exploit the different physical behaviors of our signal of interest, $XCO_2$, and confounding signals like aerosols and land surface reflectance over time. Aerosols, characterized primarily by Band 1 (0.47 μm), exhibit high-frequency, transient variations as plumes or dust events undergo rapid advection or dispersion (Kaufman et al., 1997). Simultaneously, surface reflectance follows a more deterministic, periodic behavior driven by the Bidirectional Reflectance Distribution Function (BRDF) as the SZA shifts, together with fluctuations in infrared emissivity due to surface thermal inertia (Schaaf et al., 2002). In contrast to these transient artifacts, $XCO_2$ variations within a 100-minute window are dominated by synoptic-scale transport and atmospheric advection (Parazoo et al., 2008). This temporal filtering enhances the signal-to-noise ratio (SNR) and helps the model isolate localized carbon absorption from the high-frequency noise of environmental confounders. The window was capped at 100 minutes to maintain inference viability within our $< 65^0$ constraint, ensuring consistent illumination geometry throughout the observation sequence.

## ECMWF weather reanalysis

To constrain the physical state of the atmosphere and surface, we integrated hourly meteorological data from the European Centre for Medium-Range Weather Forecasts (ECMWF) Reanalysis v5 (ERA5) (Hersbach et al., 2020). These features provide the essential thermodynamic and surface context required to de-confound the primary satellite signal, ensuring the architecture accounts for varying air mass properties and boundary layer dynamics that influence both radiative transfer and localized carbon fluxes. The extracted variables include information on 1) the Surface and Boundary Layer State: Surface pressure, 2-meter temperature and dew point, 10-meter wind components ($U$ and $V$), skin temperature, boundary layer height, and fractional coverage of high and low vegetation; 2) Atmospheric Profiles: Air temperature sampled across six vertical pressure levels (100, 200, 300, 500, 700, and 850 hPa) and specific humidity sampled across three vertical levels (300, 500, and 850 hPa) to characterize the thermodynamic structure of the troposphere and lower stratosphere; 3) Column-Integrated Quantities: Total column water vapor and total column ozone. The hourly ERA5 fields were temporally matched to the nearest GOES-East scan within a two-hour window and spatially projected onto the Advanced Baseline Imager (ABI) grid using bilinear interpolation.

## MODIS surface reflectance

To account for surface albedo—a primary confounding variable in trace gas remote sensing—we used 8-day composite surface reflectance data (Bands 1–7) from the Moderate Resolution Imaging Spectroradiometer (MODIS) Terra and Aqua products (MOD09A1.061) (Vermote, 2021), accessed via the Google Earth Engine platform (Gorelick et al., 2017). These data establish a baseline surface context, allowing the model to disentangle atmospheric variations from land-surface variability and explicitly control for albedo effects during $XCO_2$ estimation. We also explicitly passed an Normalized Difference Vegetation Index to the network, defined as (NIR - Red) / (NIR + Red), where Red and NIR are band 1 (620–670 nm ) and band 2 (841–876 nm) from MODIS. The 8-day composites were temporally matched to the GOES-East observation dates within a ±16-day window and spatially bilinearly interpolated onto the fixed GOES-16 grid.

## Total Carbon Column Observing Network soundings

To independently validate the machine learning model, we complemented the satellite-derived OCO-2 and OCO-3 retrievals with a set of 185,533 ground-based observations from the Total Carbon Column Observing Network (TCCON) GGG2020 release. Aligning with our 2024 temporal hold-out framework, we selected stations operating during that year that fell within the GOES-East field of view: Park Falls (Wisconsin, USA; Wennberg et al., 2022b), East Trout Lake (Saskatchewan, Canada; Wunch et al., 2022), Edwards (California, USA; Iraci et al., 2022), Lamont (Oklahoma, USA; Wennberg et al., 2025), and Caltech (California, USA; Wennberg et al., 2022a). These ground-based, upward-looking spectrometers provide precise, continuous column measurements, allowing us to evaluate the model's predictive skill across both diurnal and seasonal cycles at fixed geographical locations.

We evaluated the model by adapting the framework from Das et al. (2025). Specifically, we employed a spatial bounding region of 2.5° x 5.0° latitude-longitude centered on the background TCCON sites (Park Falls, East Trout Lake, and Lamont), and a constrained 0.5° x 0.5° box for urban sites (Edwards and Caltech) to mitigate complex terrain and boundary layer effects. To account for instruments with different sampling frequencies, temporal coincidence was established at two distinct scales. For the satellite baseline, OCO-2 overpass retrievals were matched against the median of TCCON measurements recorded within a ±1-hour window of the target observation, forming a daily overpass baseline. Conversely, the continuous predictive skill of DeepX$CO_2$ was evaluated at high temporal resolution using all available 10-minute scans within the local noon window (12:30–14:30 LST), matched to corresponding 10-minute TCCON medians. Although physical validations typically apply the satellite's instrument averaging kernel to the TCCON profiles to perfectly align vertical sensitivities (Das et al., 2025), we omitted this correction. Because our machine learning model was trained directly on the OCO-2 bias-corrected Level 2 XCO2 product, it implicitly maps to the satellite's native vertical sensitivity during prediction. Furthermore, applying the OCO-2 averaging kernel typically alters the TCCON column estimate by less than 0.1 ppm (Wunch et al., 2017)—a statistically negligible shift relative to the model's predictive uncertainty.

Building on this framework, the aggregate evaluation (Figure 5) uses a three-tier comparative analysis. We evaluated OCO-2 performance on successful nadir-only overpass days (N=22 overpasses in 2024), enforcing strict minimums of 100 and 15 quality-filtered soundings for OCO-2 and TCCON, respectively. This baseline was compared against the continuous 10-minute DeepX$CO_2$ performance within the local noon window. Finally, to ensure an equivalent temporal baseline for comparison, we isolated a strictly matched DeepX$CO_2$ subset, restricting the 10-minute evaluations exclusively to the specific calendar days corresponding to valid OCO-2 overpasses.

For the individual site-level evaluation (Figure 6), we compared model skill against each site’s inherent synoptic and seasonal atmospheric variability. To capture this variance and increase the paired sample size, we relaxed the sounding constraints to a minimum of 5 valid, quality-filtered observations for both the OCO-2 overpass and the corresponding TCCON window. Alongside the satellite and model residual distributions (defined as the difference between Estimates and TCCON), we calculated natural TCCON variability distributions for both the daily and 10-minute temporal scales. Both zero-centered baselines are calculated by subtracting the mean of the overpass daily medians from the respective TCCON evaluation interval. By establishing these benchmarks, we can show that predictive residuals narrower than the natural anomaly successfully track localized atmospheric volatility rather than simply regressing to the mean.

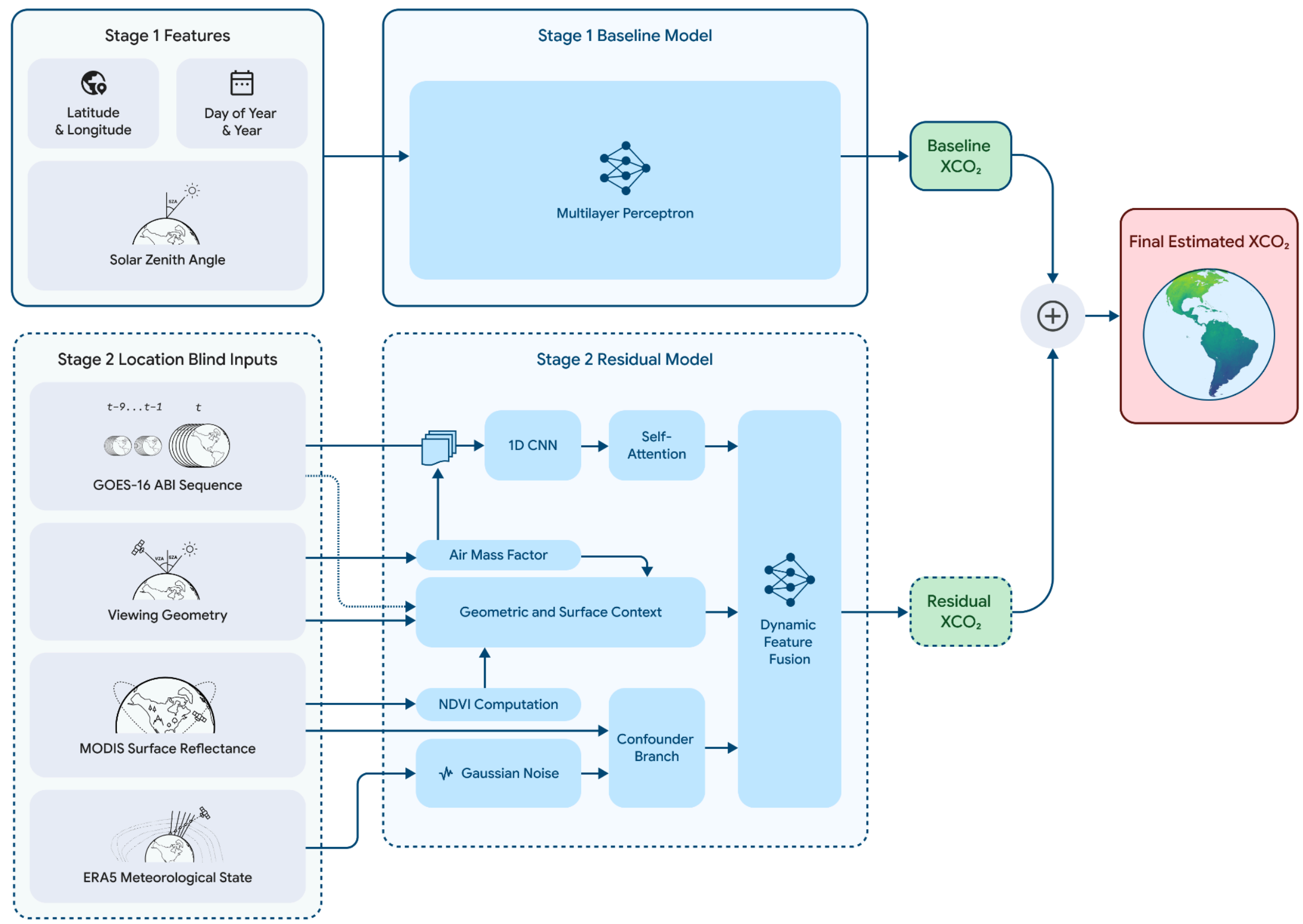

Figure 2. Schematic of the two-stage residual neural network architecture for $XCO_2$ estimation. In Stage 1 (top), a baseline Multilayer Perceptron (MLP) ingests predictable spatiotemporal features (latitude, longitude, time, and solar zenith angle) to establish a baseline $XCO_2$ prior. In Stage 2 (bottom), a deep residual network predicts the dynamic $XCO_2$ anomaly. To enforce Physics generalization and prevent location memorization, Stage 2 is explicitly blinded to static geospatial coordinates and processes dynamic inputs through specialized parallel branches. Physical Air Mass Factor (AMF) calculations are fused directly with GOES-16 ABI sequences prior to temporal processing via a 1D CNN and Self-Attention module. Concurrently, ERA5 meteorological states are subjected to a Gaussian noise bottleneck to force model reliance on spectral satellite radiances, while MODIS surface reflectance is used to explicitly compute NDVI. These features, alongside aerosol proxies and dedicated slant-path geometric corrections, converge in a deep Dynamic Feature Fusion block. The final estimated $XCO_2$ is the sum of the Stage 1 baseline prior and the learned Stage 2 dynamic residual.

## Deep learning Two-Stage framework

Carbon cycle dynamics have strong spatio-temporal regularity, characterized by an interannual growth rate, seasonal biospheric fluxes, and recurring anthropogenic emission patterns, which makes them highly predictive. To leverage this, while preventing model over-reliance on historical patterns, we designed a two-stage residual modeling framework as shown in Figure 2. This architecture was explicitly focused on decoupling predictable carbon cycle gradients from dynamic atmospheric signals, preventing the neural network from simply memorizing geospatial patterns or interpolating reanalysis data. By isolating a baseline in the first stage and letting it extract predictable patterns, the second stage is forced to retrieve meaningful physical $XCO_2$ deviations directly from the GOES-East broadband channels, the dynamic weather state and the surface reflectivity. Using these last three sets of features exclusively on the second stage lets us rigorously account for the $XCO_2$ signal as well as confounding factors—such as surface reflectance, atmospheric humidity, and complex viewing geometries—ensuring the model estimates $XCO_2$ based on fundamental remote sensing physics. Additionally, this two-stage framework enhances interpretability by explicitly distinguishing the expected seasonal baseline and large-scale geographic gradients from observed dynamic deviations. The neural network architecture was implemented and trained using the TensorFlow framework (Abadi et al., 2016).

## Stage 1 Baseline MLP

To establish a baseline for large-scale patterns, we first trained a Multilayer Perceptron (MLP) using strictly predictable spatio-temporal covariates: year, latitude, longitude, and the cosines of both the day of the year (DOY) and SZA, which we refer to as the “baseline model.” Because the model used a single-pixel context, we implemented a hemispheric phase-adjustment to the temporal features. By inverting the DOY cosine curve for the Southern Hemisphere, we mathematically synchronized 'summer' and 'winter' across latitudes, enabling the MLP to learn a unified representation of large-scale atmospheric $CO_2$ dynamics rather than relying on rigid calendar dates. Consequently, this Stage 1 model functions as a high-dimensional spatiotemporal lookup table, mapping the predictable global carbon cycle—including secular trends, latitudinal gradients, and seasonal cycles—to provide a robust baseline for the residual learning stage.

The architecture consists of two fully-connected hidden layers with 128 units each, using Rectified Linear Unit (ReLU) activations (Nair & Hinton, 2010) and an L2 regularization penalty of 0.01 to prevent overfitting (Krogh & Hertz, 1991). Input features were standardized through a normalization layer pre-fit on the training data distribution (LeCun et al., 2012). We optimized the model using Huber Loss (delta = 1.2) (Huber, 1964) to ensure that outlier measurements from OCO-2 and OCO-3 did not disproportionately skew the baseline parameters. This loss function maintains the stability of Mean Squared Error (MSE) for small residuals while transitioning to Mean Absolute Error (MAE) for larger deviations, effectively preventing the model from over-adjusting to anomalous pixels. The model was trained for 75 epochs (using a batch size of 512 for 976 steps) with an initial learning rate of 0.001. To ensure convergence, we paired this with a reduce-on-plateau decay schedule (Bengio, 2012; Goodfellow et al., 2016).

## Stage-2 Multi-Branch Residual Architecture

Following the convergence of the baseline model, its weights were frozen to preserve the learned historical patterns and ensure a stable reference. We then trained a second, high-capacity residual network to predict the deviations between the observed OCO-2 and OCO-3 soundings and the baseline predictions. We henceforth refer to this stage-2 model as the “residual model,” and we use the name “Deep$XCO_2$” to refer to the sum of the predictions from the stage 1 and stage 2 models. To ensure the model prioritized physical signal extraction over geospatial memorization, this residual stage was explicitly denied access to coordinates (latitude, longitude) or the year. Instead, the model operated exclusively on dynamic, observation-time data streams: a 100-minute GOES-East temporal sequence (comprising the concurrent scan and the nine preceding 10-minute intervals), 8-day average MODIS surface reflectance, and hourly ERA5 meteorological states. This residual model processes these inputs through three specialized parallel branches: 1) a GOES-East temporal branch designed to capture atmospheric and diurnal dynamics, 2) an ERA5 and MODIS confounder-mitigation branch to account for atmospheric state and surface reflectance, and 3) a physical Air Mass Factor (AMF) correction branch to normalize for varying optical path lengths.

The ERA5 data is highly informative of atmospheric $CO_2$ variability because synoptic-scale meteorology (including frontal passages and boundary layer dynamics) is a primary driver of non-local $XCO_2$ variations (Parazoo et al., 2008). While estimating this covariance is essential for de-confounding the GOES-East channels, over-reliance on meteorological information poses a risk of the network becoming a 'weather-based emulator' that fails to exploit the spectral signal and high-spatiotemporal resolution unique to GOES-East. To mitigate this, we implemented an information bottleneck by injecting Gaussian noise into the ERA5 features during training with a standard deviation (STD) of 0.15, this was a 15% signal-to-noise perturbation relative to the normalized feature’s STD of 1. Beyond restricting mutual information, this noise injection acts as a robust regularizer that accounts for the inherent representation errors, sub-grid scale parameterization uncertainties, and structural biases present in reanalysis products, preventing the model from overfitting to imperfect meteorological fields. This intentional degradation of meteorological reliability forces the network to prioritize the subtle gradients within the GOES-East radiances while training, which provide a weaker but more direct causal signal of localized $CO_2$ concentration. Concurrently, this branch processes raw MODIS Red and Near-Infrared bands and computes an explicit Normalized Difference Vegetation Index (NDVI), providing the network with a vital proxy for localized biospheric flux potential.

While our training targets rely exclusively on nadir observations from low-Earth orbit sensors (OCO-2 and OCO-3), GOES-East operates from a geostationary perspective. Consequently, pixels distant from the sub-satellite point exhibit substantial geometric distortions that alter the optical path length of the measured radiation. To explicitly normalize for these variations, this branch calculates the components of the physical Air Mass Factor (AMF) as a function of the SZA and the satellite viewing zenith angle (VZA), defined by their respective secants: 1/cos(SZA) and 1/cos(VZA). These standardized physical AMF terms are injected directly into both the dedicated geometric processing layers and the GOES-East temporal branch, equipping the network with the explicit physical boundaries required to map non-linear slant-path corrections and cleanly isolate true atmospheric tracer absorption from geometric modulations.

The third parallel branch ingests the GOES-East radiance data as a 16-dimensional time series to capture fine-scale atmospheric dynamics. This branch executes three cascading operations to process the temporal input: a 1-dimensional Convolutional Neural Network (1D CNN) that scans across the temporal dimension to extract localized, high-frequency feature representations; an 8-head self-attention mechanism (Vaswani et al., 2017) that evaluates the entire 10-scan sequence to model long-range temporal dependencies; and a global statistical pooling layer. The self-attention block explicitly allows the network to assign higher weights to optimal observation moments, dynamically isolating clear-sky intervals and maximizing the signal-to-noise ratio of the subtle $XCO_2$ absorption features within the 100-minute window. To condense this processed sequence into a single, dense feature vector, global average and global max pooling outputs are concatenated at the final stage of the branch (Lin et al., 2013). In practice, this dual-pooling strategy outperformed fully connected linear layers when applied to the attention outputs. By extracting non-parametric summary statistics directly through simultaneous average and max pooling operations, the model preserves distribution characteristics without introducing excessive parameters and complexity. These summary statistics capture both the steady-state baselines and the transient extremes of the temporal distributions, letting the downstream layers track $CO_2$-sensitive spectral band behavior over time and associated dynamic confounders, such as surface reflectance, atmospheric humidity, and aerosols.

## Training Framework

The residual model was optimized using Huber loss to ensure robust parameter estimation against observational anomalies. To promote global generalization and mitigate overfitting within the high-capacity architecture, we used several regularization techniques: an elevated dropout rate (35%) and L2 weight decay penalties across nearly all dense and convolutional layers. Additionally, we used Swish activation functions in the deep state branches (Ramachandran et al., 2017). The smooth, non-monotonic properties of the Swish function help with vanishing gradient issues during backpropagation, providing a critical architectural adjustment for preserving the signal flow of attenuated, subtle features—such as the marginal trace-gas absorption signatures present within the 13.3 μm channel of GOES. Finally, we implemented a strategic sample weighting scheme to capture diurnal carbon dynamics. While OCO-2 operates in a sun-synchronous orbit with a fixed 1:30 pm local overpass time, OCO-3 is installed on the International Space Station, providing non-sun-synchronous observations across various times of day. To prevent the model from overfitting to afternoon atmospheric physics, OCO-3 training examples were upweighted by a factor of 2x relative to OCO-2 data, forcing the network to generalize across the full diurnal cycle.

To systematically explore the architectural search space and fine-tune hyperparameter dynamics, we utilized the automated code generation framework introduced by Aygün et al. (2025). This advanced system couples Large Language Models (LLMs) for iterative, syntax-valid code synthesis with a Monte Carlo Tree Search (MCTS) to sample and optimize target performance surfaces. Using Gemini 3 Flash (Google, 2025) as the generative backbone, the framework navigated complex structural choices using the Predictor Upper Confidence Bound (PUCB) heuristic (Silver et al., 2016). The framework did not operate as a blind search; instead, it functioned as an "expert"-guided Automated Machine Learning system (Hutter et al., 2019) which has a physics “understanding” of the underlying features and can come up with new ones through feature engineering. We supplied the pipeline with our foundational domain-specific constraints, including the Stage 1 frozen baseline, the physical Air Mass Factor (AMF) geometry path, and the ERA5 information bottleneck noise regularizer. Acted upon by these strict physical boundaries, the MCTS loop accelerated our development cycle by rapidly iterating on network configurations (such as layer widths, Swish activation patterns, and 8-head temporal attention schedules). This interaction between human physical understanding and machine optimization successfully stabilized the deep regressions.

## Signal Attribution

We conducted an ablation study to quantify the contribution of individual data sources to the model's predictive accuracy (Table 1). We evaluated four targeted exclusion scenarios: excluding all GOES-East channel time series (a.k.a. “No GOES”), removing the weather data (a.k.a. “No ERA5”), removing surface albedo context (a.k.a. “No MODIS”), and removing the GOES-East channels most sensitive to $CO_2$ (a.k.a. “No Band 5 & 16”). To ensure a rigorous comparison across the exact same architecture, the ablations were not performed by altering the model's input dimensions or layer structures, which could inadvertently penalize the model by reducing its overall parameter capacity. Instead, the targeted input tensors were systematically replaced by zeros within the computational graph during both training and inference. This zero-masking approach successfully destroyed the targeted physical signals while preserving identical network architecture, data flow, and model capacity across all experiments. The ablated networks were then evaluated against both the full feature set and the baseline models to assess their spatial generalization (via strictly OCO-2 collocations, as opposed to the joint OCO-2/OCO-3 dataset used in the primary evaluation) and their sensitivity to local boundary layer dynamics (via TCCON observations).

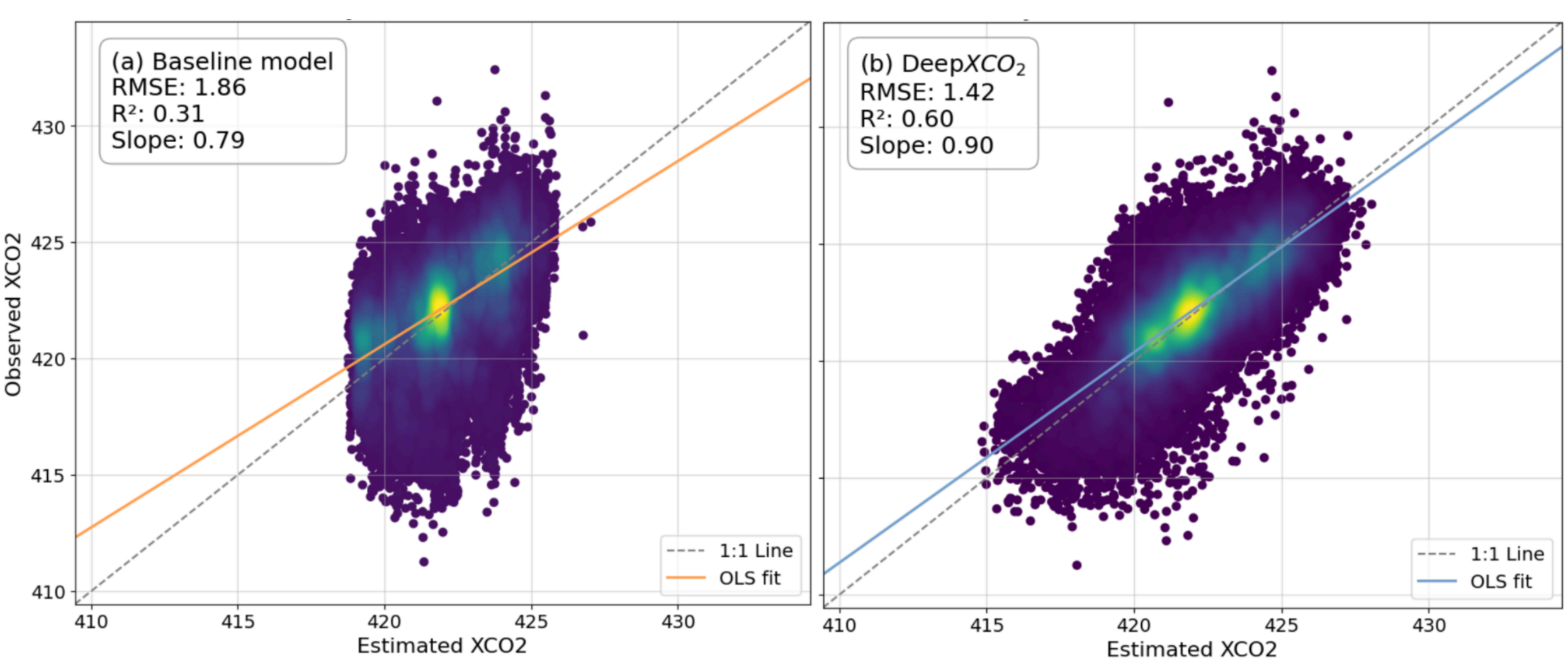

Figure 3: Density scatter plots comparing estimated versus observed OCO-2 and OCO-3 $XCO_2$ concentrations for the Stage-1 baseline model (left) and the Stage-2 DeepX$CO_2$ (right). The color gradient denotes local data point density, with warmer colors (yellow/green) indicating regions of higher data concentration. In each panel, the dashed gray line represents the 1:1 line of perfect agreement, while the solid line indicates the Ordinary Least Squares (OLS) regression fit. Inset text boxes display the overall evaluation metrics, including micro Root Mean Square Error (RMSE), the coefficient of determination ($R^2$),

and the regression slope. DeepXCO$_2$ shows superior skill, based on the lower RMSE (1.42 vs. 1.86 ppm), higher $R^2$ (0.60 vs. 0.31), and a regression slope closer to one (0.90 vs. 0.79).

# Results

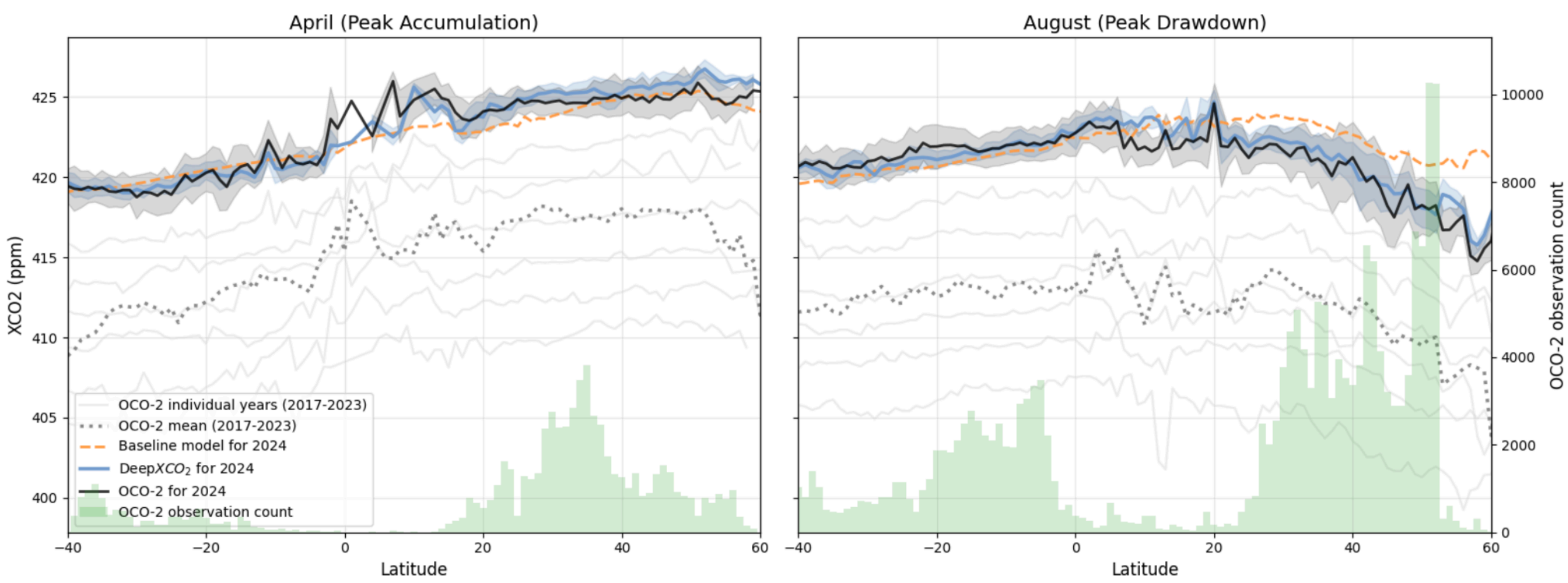

Figure 4: Latitudinal gradients of $XCO_2$ during the Northern Hemisphere's peak accumulation (April, left) and peak biological drawdown (August, right). The solid black line represents the 2024 OCO-2 ground truth observations, while the solid blue line denotes the DeepXCO$_2$ predictions. The dashed orange line represents the static Stage-1 baseline. Individual training set years (faint gray lines) and the training set OCO-2 mean (dotted gray line) are shown for context. Shaded bands indicate ±1 standard deviation, representing the longitudinal variability within each 1° latitude bin. The background histogram (light green, right y-axis) displays the total count of valid OCO-2 observations available at each latitude.

## Evaluation against held-out OCO-2 and OCO-3 observations

Evaluation of $XCO_2$ estimates obtained from DeepXCO$_2$ for 2024 against coincident OCO-2 and OCO-3 nadir observations yields a root mean square error (RMSE) of 1.42 ppm and a coefficient of determination ($R^2$) of 0.60 (Figure 3b). The baseline model, which does not include GOES-East observations, ERA5 meteorological data, or MODIS surface reflectance yields a higher RMSE of 1.86 ppm and lower $R^2$ of 0.31 (Figure 3a), confirming that the performance of the full model is enabled by the addition of relevant remote sensing information, rather than simply being the result of the model memorizing historical spatial and temporal variability in $XCO_2$. Naturally, the RMSE based on DeepXCO$_2$ is higher than the precision ceiling accuracy of OCO-2 land nadir observations (~ 0.85 ppm, Das et al. 2025); this is expected given the very broad spectral range of its channels (e.g., 1580 to 1640 nm for channel 5 and 13000 to 13600 for channel 16).

The inferred latitudinal gradient based on DeepXCO$_2$ at times and locations coincident with OCO-2 observations for 2024 further confirms that the spatial gradients inferred from DeepXCO$_2$ are realistic (Figure 4). The gradient inferred based on DeepXCO$_2$ also more closely resembles the latitudinal gradient for the evaluation year (i.e., 2024) relative to that for other years (2017-2023), confirming that DeepXCO$_2$ is able to correctly capture interannual variability.

## Evaluation against TCCON observations

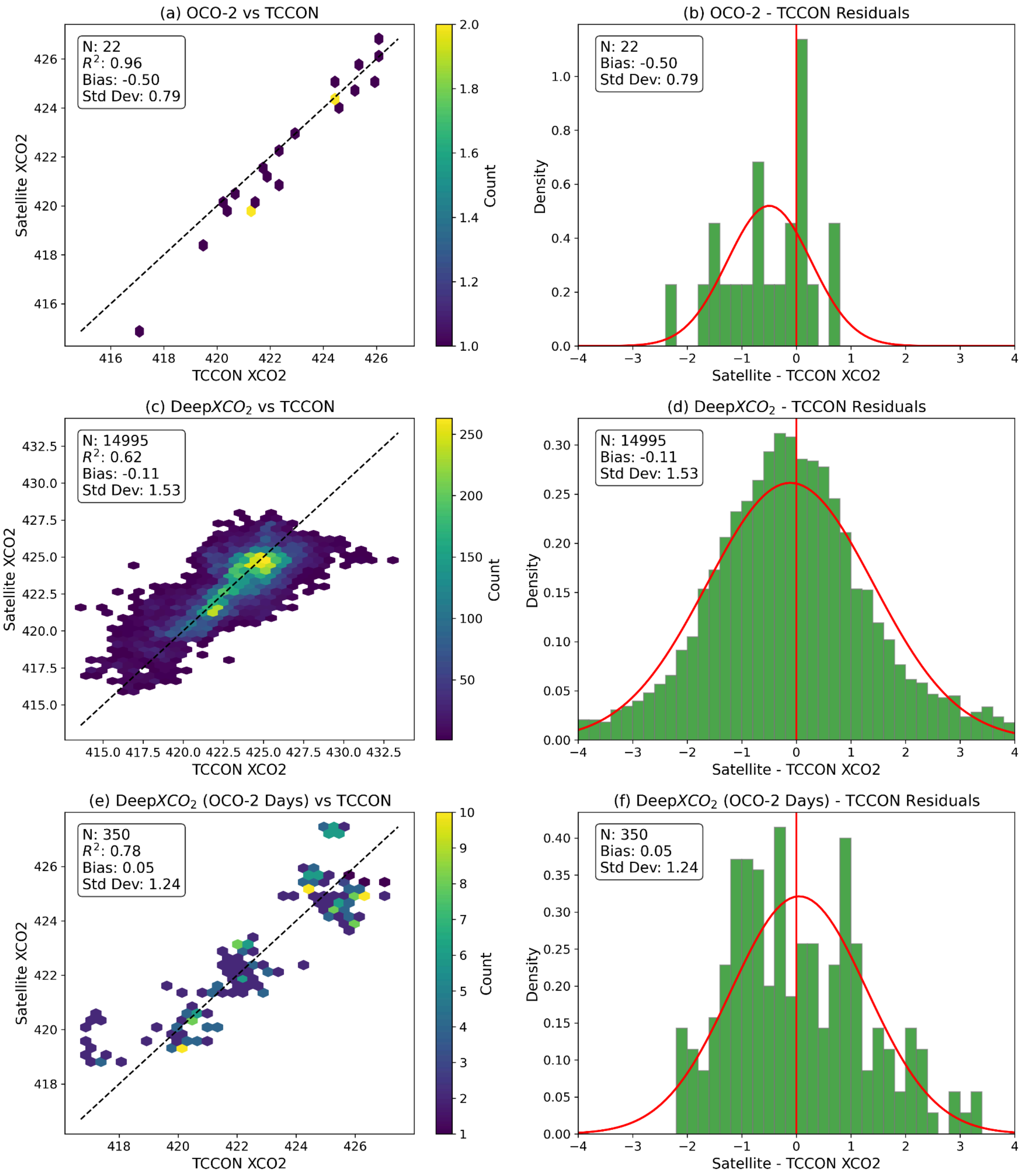

Figure 5: Evaluation of OCO-2 retrievals and DeepXCO$_2$ estimates against ground-based TCCON observations. Left column shows hexagonal bin density scatter plots against TCCON data with a 1:1

reference line. Right column shows error distribution histograms of the residuals (Delta $XCO_2$ = Estimate - TCCON) with overlaid Gaussian density fits and zero-bias reference lines. Relevant statistical metrics are displayed in each panel. (a, b) OCO-2 vs. TCCON baseline satellite performance on successful overpass days (N = 22), matched to the TCCON median within a 1-hour window. (c, d) $DeepXCO_2$ vs. TCCON model estimates evaluated continuously across all available 10-minute intervals during the local noon window (12:30–14:30 LST; N = 14,995). (e, f) $DeepXCO_2$ (OCO-2 Days) vs. TCCON temporally matched subset (N = 350) isolating 10-minute model evaluations exclusively on calendar days featuring a successful OCO-2 overpass to establish an equivalent baseline. Restricting the 10-minute model evaluation to these satellite-observed days establishes an equivalent temporal baseline. This demonstrates consistency in bias and variance, confirming that the model maintains stable predictive fidelity across both highly sampled and unobserved periods.

We further evaluate $DeepXCO_2$ against observations from five TCCON sites located within the GOES-East field of view: Park Falls, Lamont, Edwards, Caltech, and East Trout Lake. For ease of evaluation, we follow a similar approach to that implemented for the latest OCO-2 retrieval algorithm (Das et al., 2025; see Methods). Specifically, we reproduce the analytical frameworks of Figures 3 (top panels) and 8 from that study to assess how well $DeepXCO_2$ captures observed site-level variability. However, to account for the continuous temporal resolution of our geostationary model, we adapt the matching criteria. Whereas OCO-2 provides at most a single ~1:30 PM local overpass within the region defined by the spatial coincidence criteria, we evaluate $DeepXCO_2$ continuously. Over a two-hour window ranging from 12:30 PM to 2:30 PM local time, we compare every 10-minute scan estimate within the target region directly against its corresponding 10-minute TCCON median.

Under these conditions, we find a standard deviation of 0.79 ppm for OCO-2 observations versus 1.53 ppm for $DeepXCO_2$ (Figure 5).  The $R^2$ is 0.96 for OCO-2 and 0.62 for $DeepXCO_2$. Observations from the two instruments show low bias (-0.50 ppm for OCO-2 and -0.11 ppm for $DeepXCO_2$). Note that the values reported here for OCO-2 land nadir are not identical to those reported in Das et al. (2025), because we do not use glint mode soundings, and both the region and time span of the intercomparison are different.

Examining the performance of the model broken out by individual TCCON sites confirms the strong performance of $DeepXCO_2$. Figure 6 shows residuals between OCO-2 observations and coincident TCCON observations, as well as residuals between $DeepXCO_2$ and coincident TCCON observations. The full observed variability at the TCCON sites is shown for comparison. The fact that estimates from $DeepXCO_2$ show only mild degradation relative to OCO-2 is notable given that OCO-2 is a dedicated hyperspectral instrument while GOES is a multispectral instrument with only 16 bands.

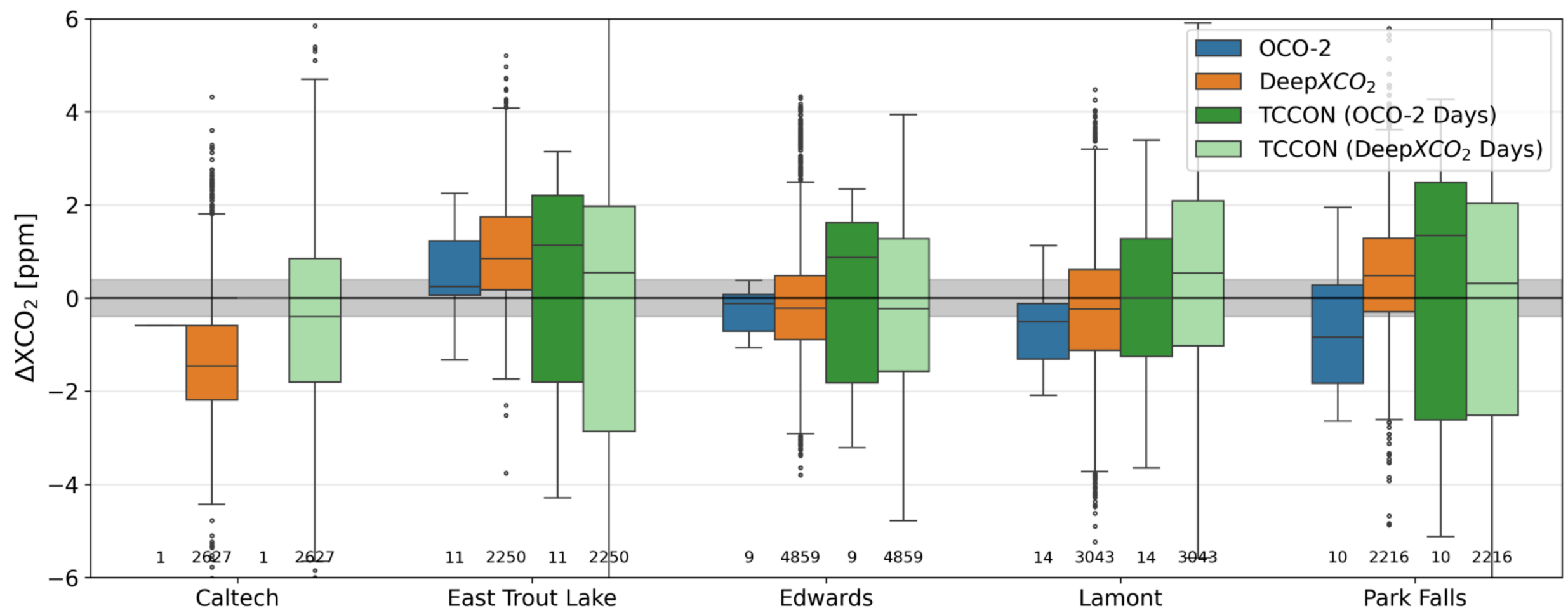

Figure 6. Site-level distribution of validation residuals and natural atmospheric variability under relaxed sample-size constraints (minimum of 5 observations per window). Box plots display the distribution of Delta XCO2 residuals across five evaluating TCCON stations, with sample sizes for each configuration shown above the x-axis and a shaded grey band highlighting a ±0.4 ppm reference region. Residuals are plotted for OCO-2 daily overpasses matched to a ±1-hour TCCON median (Blue) and continuous Deep$XCO_2$ 10-minute estimates matched within the local noon window (Orange). Zero-centered empirical baselines illustrate natural TCCON atmospheric variability restricted to successful OCO-2 overpass days at a daily scale (Dark Green) and across all valid Deep$XCO_2$ local noon intervals at a 10-minute scale (Light Green). These TCCON distributions serve as empirical environmental baselines; predictive residuals that remain narrower than the natural anomaly indicate that the respective models successfully track atmospheric volatility rather than simply regressing to the mean.

## Ablation study

The ablation study (Table 1) confirms that GOES-East observations are the primary source of information for the model. Removing the entire GOES time series degrades predictive accuracy more (dropping OCO-2 $R^2$ to 0.60 and TCCON $R^2$ to 0.27) than removing either the ERA5 meteorological data or MODIS features. Interestingly, dropping only the $CO_2$-sensitive GOES channels (Bands 5 and 16) noticeably reduces performance at the TCCON towers (dropping $R^2$ from the 0.43 baseline to 0.33), but only marginally affects the OCO-2 baseline. This resilience in OCO-2 prediction accuracy is possibly due to the high covariability among the remaining longwave infrared window channels, which continue to provide sufficient redundant information for estimating the bulk total column from space. This redundancy is not limited to the GOES signal; the minimal drop in OCO-2 $R^2$ observed when dropping either ERA5 or MODIS highlights that the model's multiple branches synthesize complementary signals distributed across all three data streams, preventing failure when one source is unavailable.

| Ablation | OCO-2 | | TCCON | |
|---|---|---|---|---|
| | RMSE | $R^2$ | RMSE | $R^2$ |
| All Features | 1.37 | 0.64 | 1.65 | 0.43 |
| No Band 5&16 | 1.38 | 0.64 | 1.78 | 0.33 |
| No GOES | 1.45 | 0.60 | 1.81 | 0.27 |
| No ERA5 | 1.37 | 0.64 | 1.80 | 0.31 |
| No MODIS | 1.39 | 0.63 | 1.79 | 0.34 |
| Baseline | 1.76 | 0.41 | 2.26 | 0.02 |

Table 1. Ablation study of DeepX$CO_2$ predictions. Performance is evaluated using Root Mean Square Error (RMSE, in ppm) and R2 across OCO-2 and TCCON observations (filtered for Solar Zenith Angle < 65$^o$).

## Urban observation case study

Case studies over urban environments demonstrate the benefit of the high spatial and temporal density of GOES observations. Figure 7 presents snapshots of Los Angeles (October 18, 2024), Chicago (October 6, 2024) with observations averaged across 12:30 pm to 2:30 pm local time. The urban X$CO_2$ enhancement is clearly visible over the urban areas. While the individual observations from DeepX$CO_2$ are not expected to be as accurate as those from OCO-2, the ability to have full coverage every 10 minutes means that in the aggregate this model is providing unprecedented information at high spatial resolution. For comparison, Figure 7 also shows the available OCO-2 nadir observations on those same days.

Interestingly, these case studies also illustrate the fact that, because the model is trained on OCO-2 and OCO-3 data, it will also inherit (i.e., learn) some of the remaining biases inherent to those data. One example of this is visible in the Chicago case study, where river networks appear in blue (i.e., lower inferred X$CO_2$) and this same effect is visible in the single OCO-2 overpass on that same day. Some of these biases in the OCO-2 and OCO-3 data are well known and documented (e.g., Das et al., 2025).

Examples of inferences over Chicago are compared in Figure 8 to observations from occasional October OCO-3 Snapshot Area Maps (SAMs) and to nitrogen dioxide ($NO_2$) observations from the Sentinel-5 Precursor TROPOMI instrument (van Geffen et al., 2020; Veefkind et al., 2012). Spatial patterns of atmospheric $NO_2$ concentrations have previously been used in the assessment of SAMs because of the strong spatial co-variability between $CO_2$ and $NO_2$ emissions in urban regions. These comparisons confirm that the urban X$CO_2$ estimated from DeepX$CO_2$ is realistic within the context of other available observations.

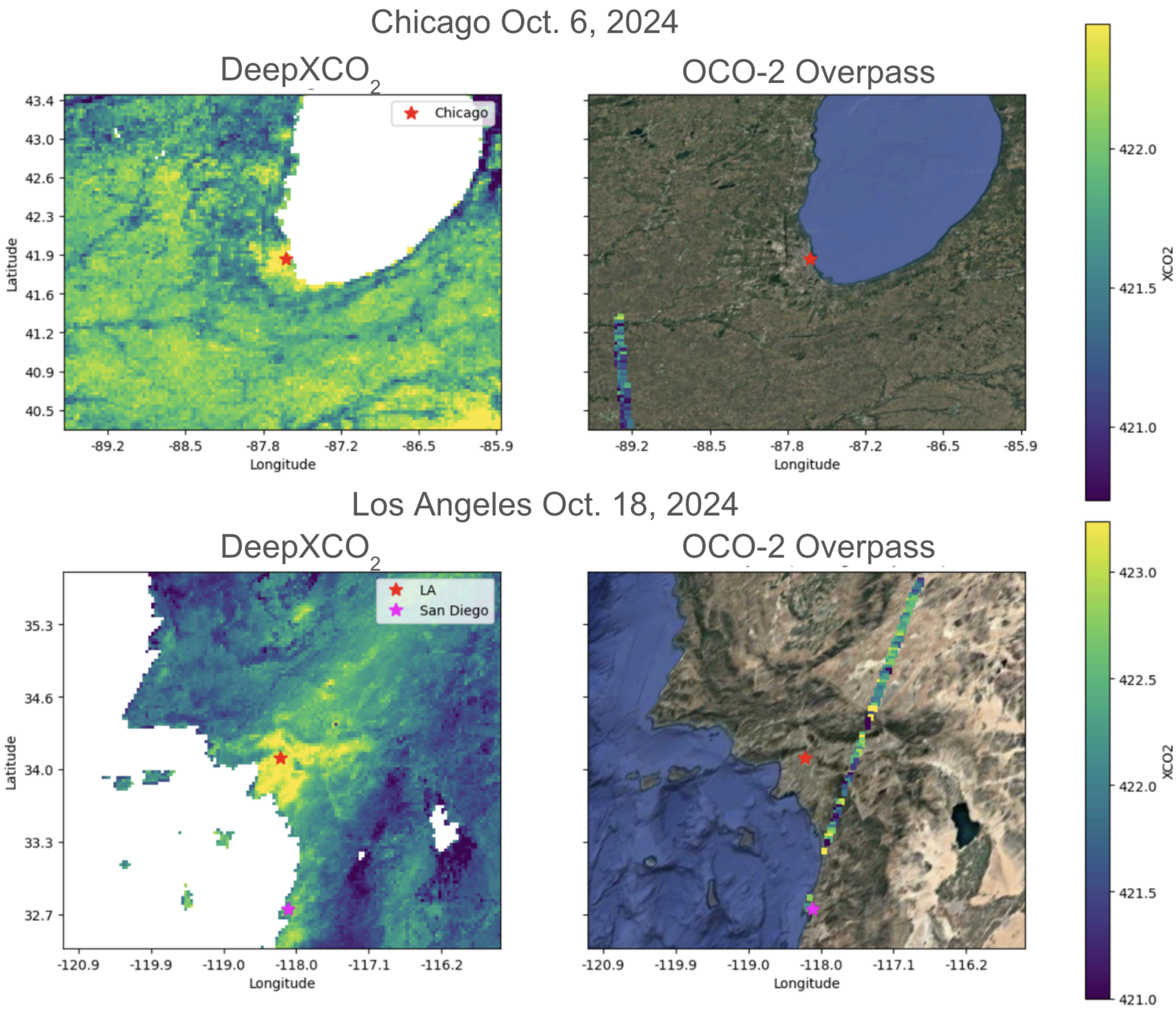

Figure 7: Urban case studies for Los Angeles and Chicago. Mid-afternoon atmospheric column-averaged carbon dioxide ($XCO_2$) over Chicago on October 6, 2024 (top row) and Los Angeles on October 18, 2024 (bottom row). Left panels: $XCO_2$ estimates derived from DeepXCO$_2$. The maps represent an average of consecutive scans collected between 12:30 pm and 14:30 pm local time to capture representative mid-afternoon conditions. Right panels: Coincident measurements from the OCO-2 overlaid on a satellite imagery basemap for geographical context. To facilitate direct visual comparison, OCO-2 data points are gridded to the native GOES spatial resolution (~2 km). Red stars indicate the primary city centers; in the Los Angeles panels, a pink star marks San Diego for regional context. Color bars denote column-averaged $CO_2$ (ppm).

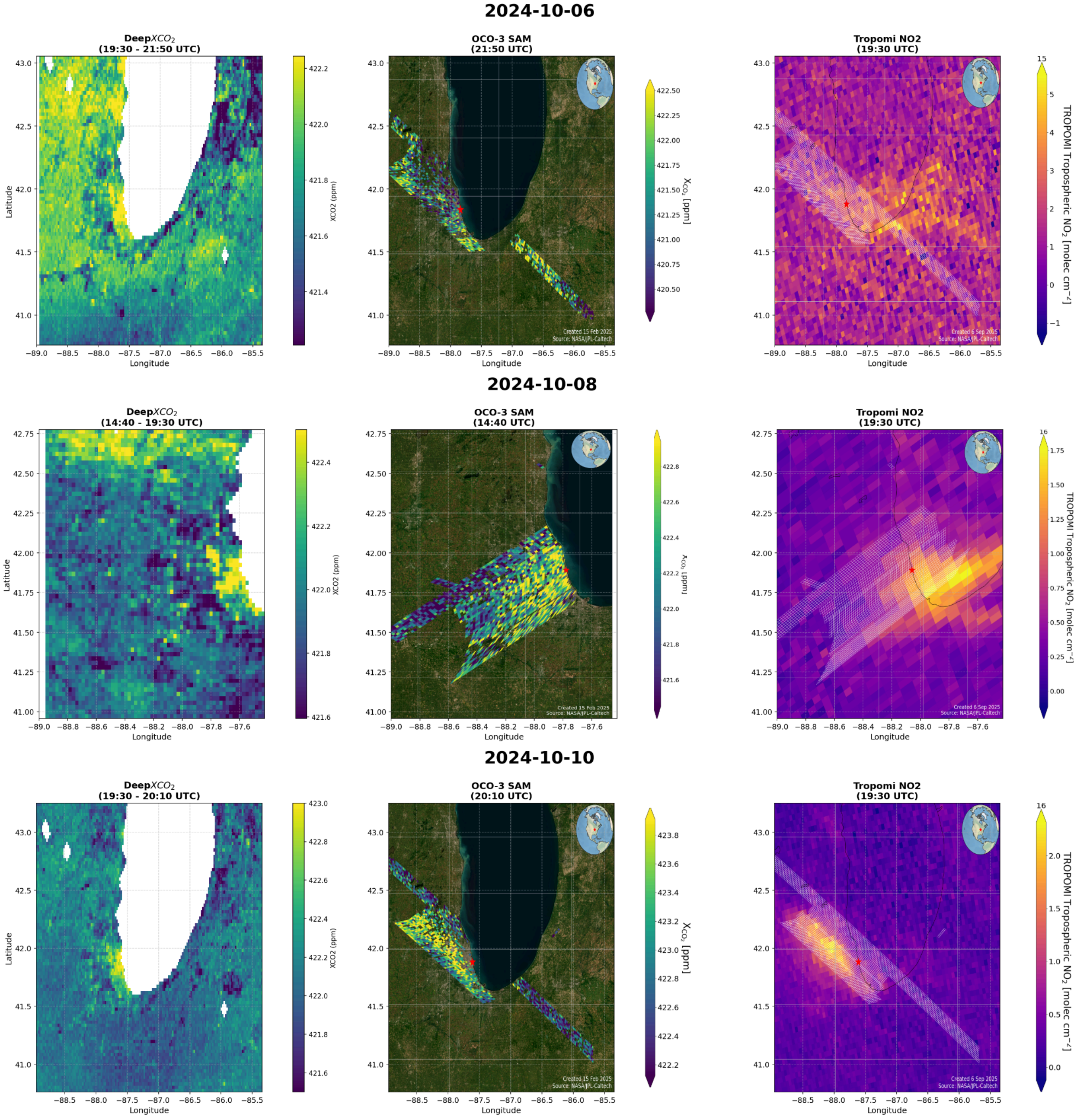

Figure 8. Spatiotemporal comparison of anthropogenic plume enhancements across the Greater Chicago Metropolitan Area during October 2024. Multi-instrument observations map fine-scale atmospheric structures across three distinct October overpass events: (a) October 6, 2024 (sID: 39342), (b) October 8, 2024 (sID: 39405), and (c) October 10, 2024 (sID: 39631). For each target date, the comparative multi-panel composites display: Left panels: DeepXCO$_2$ estimates. To account for transport dynamics between polar-orbiting sensor overpasses, the images represent a temporal average of all valid intermediate model scans strictly bounded by the primary target instrument acquisition times (noted

above each panel). Center panels (OCO-3 SAM Overpass): Spaceborne $XCO_2$ retrievals collected via the OCO-3 Snapshot Area Mode (SAM), visualizing highly localized downwind plume structures and fine-scale urban enhancements. Right panels (TROPOMI $NO_2$ Columns): Tropospheric nitrogen dioxide ($NO_2$) vertical column densities observed concurrently by the Sentinel-5 Precursor Tropospheric Monitoring Instrument, providing an independent atmospheric tracer for validating co-emitted combustion signatures.

## Agricultural region case study

Another benefit of leveraging GOES observations is the ability to observe throughout daylight hours. To illustrate this, Figure 9 presents $XCO_2$ estimated by DeepXCO$_2$, averaged over the region of the United States corn belt. The model is able to capture a regional-scale 2 ppm drawdown in $XCO_2$ during the peak of the growing season in July from mid-morning to early afternoon. In February, conversely, the model captures a higher background and no substantial diurnal variability during the dormant season. This example demonstrates that the model is able to capture realistic fine-scale temporal variability which is missed by sun-synchronous instruments, in addition to realistic spatial variability.

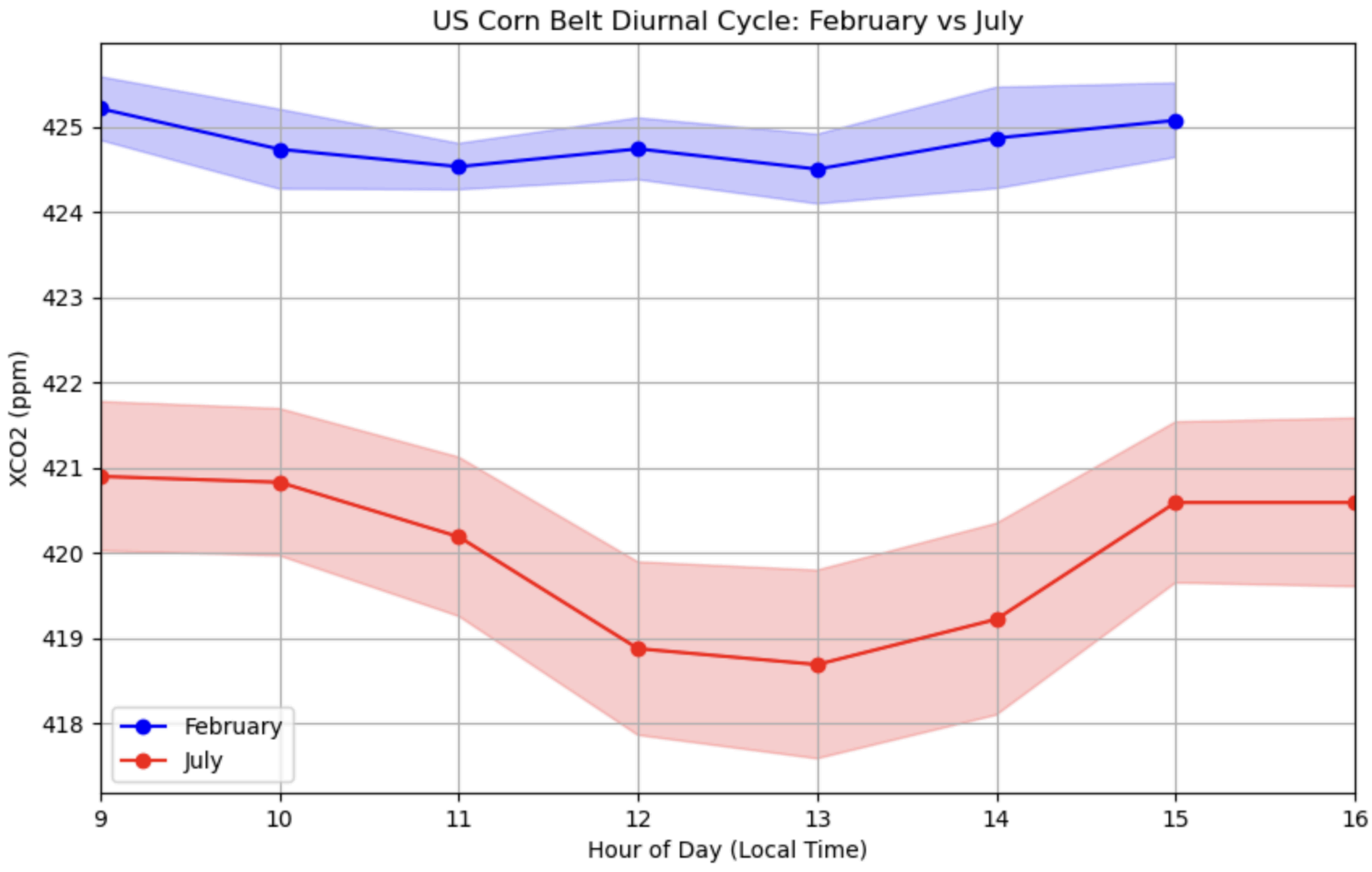

Figure 9: Diurnal cycle of XCO2 over US Corn belt for peak growing season versus winter. Red and blue lines present the diurnal cycle of XCO2 averaged over the region with latitude of 40° N to 44° N and longitude of 96° W to 88° W for July and February, respectively.

## Discussion

The current network of space-based instruments for tracking global $CO_2$ is sparse and fragile. Despite the first satellite designed specifically for making $CO_2$ observations from space (i.e., GOSAT) having been launched over 15 years ago, we still have only a very limited set of instruments that provide fully open data; these include GOSAT, GOSAT2, OCO-2 and OCO-3. In addition to how sparse these observations continue to be, observational continuity is also an ongoing challenge. For example, long-term commitments of public funding for the operation of existing missions are increasingly uncertain. Space-based observations also carry inherent risks, as demonstrated by the launch failure of the first OCO satellite in 2009 and the in-orbit failure of MethaneSat in 2025, which complicates continuity planning. In addition, funding to support data continuity is limited. Finally, there is an inherent tension between the need for data interoperability and the desire to design instruments that best support cutting edge research. These challenges are not unique to $CO_2$ monitoring and are reflective of many areas of Earth system science (KISS Continuity Study Team, 2024).

While dedicated instruments represent the gold standard for monitoring $CO_2$ from space, leveraging existing instruments originally designed for other applications presents an opportunity to address some of these challenges. $CO_2$ estimates based on observations from such instruments inherently lack the spectral sensitivity and precision of dedicated missions. However, the high spatial and temporal density of their observations, especially in the case of geostationary satellites regularly launched in support of weather forecasting, potentially presents an opportunity for a step-change in our ability to monitor global $CO_2$.

Artificial intelligence presents an opportunity to develop deep learning models that can help extract atmospheric signals from broadband multispectral data, but the development of these models is predicated on the availability of large quantities of highly accurate observations that serve as labels for training models and benchmarks for model evaluation. In the case of $XCO_2$, existing dedicated missions such as OCO-2 and OCO-3, as well as observations from the TCCON network, serve this role.

While training a machine learning model on OCO-2 and OCO-3 data leverages their high-fidelity spectral physics, it inherently means the predictions will inherit any biases remaining in those operational retrieval algorithms. We showed that $DeepXCO_2$ presented here is able to capture realistic $XCO_2$ variability when compared to independent observations from TCCON. Indeed we observed that in some cases $DeepXCO_2$ also “learned” some of the biases remaining in the OCO-2 and OCO-3 observations used in the model training. One example of this was shown in Figure 7, where inferences from both $DeepXCO_2$ and the coincident OCO-2 overpass showed local negative biases over small inland water bodies such as rivers. The propagation of these original biases in the OCO-2 and OCO-3 retrieval algorithms can in some cases make it difficult to interpret which spatial and temporal features in the $DeepXCO_2$ predictions are (1) true atmospheric XCO2 variability, (2) inherited biases from the OCO-2 and OCO-3 training data, or (3) unique artifacts specific to $DeepXCO_2$. Therefore, some features related to surface reflectance and topography, for example, will need to be explored further as part of future studies. Interestingly, this exploration could not only lead to improvements to the GOES-based model but could also inform further improvements to the OCO-2 and OCO-3 retrieval algorithms by elucidating biases through spatially- and temporally-dense analyses.

Overall, as the approach presented here demonstrates for GOES East, observations from existing instruments contain a substantial amount of information about atmospheric $CO_2$, especially when coupled with other relevant sources of information such as meteorological information and land surface properties. These new observations have the potential to begin to fill long-standing observational gaps by enabling spatially-dense observations, by providing information about $CO_2$ variability across a larger portion of the diurnal cycle, and by compensating for low data yields in cloudy regions through high frequency revisits. The case studies presented here for urban regions and for the US corn belt are examples of such opportunities.

Critically, the ability to extract complementary information about atmospheric $CO_2$ from existing space-based instruments is not a substitute for dedicated missions. This is both because the ability to use a wider array of existing instruments is predicated on the availability of high quality observations for model training and benchmarking, and because the accuracy of estimates from this broader set of instruments will never rival that of observations from hyperspectral instruments targeting $CO_2$. The use of existing instruments such as GOES is therefore a promising approach to augment, but not to substitute, the observational backbone provided by dedicated missions.

## Acknowledgments

The authors would like to gratefully acknowledge Abhishek Chatterjee, Chris O'Dell, Daniel Lindsey, Sourish Basu, Renee Johnston, Kevin McCloskey, Joe Ng, Vishal Batchu, Kirk Draheim, Rob von Behren, Nita Goyal, Carl Elkin, Steve Greenberg, and Karin Tuxen-Bettman for their insightful guidance and support. The OCO-2 and OCO-3 data were produced by the OCO-2 and OCO-3 projects at the Jet Propulsion Laboratory, California Institute of Technology, and were obtained from the NASA Goddard Earth Sciences Data and Information Services Center (GES DISC). TCCON data were obtained from the TCCON Data Archive hosted by CaltechDATA (https://tccondata.org); we acknowledge funding to support TCCON from NASA. This study contains modified Copernicus Climate Change Service information [2024] for the ERA5 meteorological reanalysis. Neither the European Commission nor ECMWF is responsible for any use that may be made of the Copernicus information or data it contains. MODIS surface reflectance data (MOD09A1.061) were retrieved from the NASA EOSDIS Land Processes Distributed Active Archive Center (LP DAAC) at the USGS Earth Resources Observation and Science (EROS) Center. Finally, this work contains modified Copernicus Sentinel data [2024] processed by ESA for the TROPOMI $NO_2$ column observations. The GOES-16 ABI data were accessed via the Google Cloud Public Datasets Program, provided through the NOAA Open Data Dissemination (NODD) Program.